\documentclass[
superscriptaddress,dvips
prb,twocolumn
]{revtex4-1}

\usepackage{graphicx,amsmath,amssymb,subfigure}
\usepackage{dcolumn}
\usepackage{bm}
\usepackage{color}

\begin{document}

\preprint{APS/123-QED}

\title{Spin wave theory study of neutron intensity, magnetic field, and anisotropy of Type IIA FCC antiferromagnet}

\author{Trinanjan Datta}
\affiliation{Department of Chemistry and Physics, Augusta State University, Augusta, GA 30904}
\affiliation{State Key Laboratory of Optoelectronic Materials and Technologies,\\
School of Physics and Engineering, Sun Yat-sen University, Guangzhou 510275, China}
\author{Dao-Xin Yao}
\email[Corresponding author:]{yaodaox@mail.sysu.edu.cn}
\affiliation{State Key Laboratory of Optoelectronic Materials and Technologies,\\
School of Physics and Engineering, Sun Yat-sen University, Guangzhou 510275, China}

\date{\today}

\begin{abstract} 
We study the spin dynamics in a 3D quantum antiferromagnet on a
face-centered cubic (FCC) lattice. The effects of magnetic field, single-ion anisotropy, and biquadratic interactions are investigated using linear spin wave theory with spins in a canted basis about the Type IIA FCC antiferromagnetic ground state structure which is known to be stable. We calculate the expected finite frequency neutron scattering intensity and give
qualitative criteria for typical FCC materials MnO and CoO. The magnetization
reduction due to quantum zero point fluctuations is also analyzed.

\begin{description}
\item[PACS numbers] 75.30.Ds, 75.10.Jm, 75.50.Ee
\end{description}
\end{abstract}
\maketitle

\section{Introduction}
Frustration plays an important role in the conceptual understanding of the physical properties of novel magnetic materials ~\cite{diep}. Frustration can arise either due to the underlying geometry as in a triangular
lattice ~\cite{weihong992D,zhit962D} case or due to competing
interactions as in the quantum spin-$1/2$ Heisenberg antiferromagnet (AF) on a square
lattice~\cite{tassi75gen,dot942D,valeri992D,singh992D,watabe91,mam06}. There are several examples of
frustrated magnetic materials: spinels~\cite{tristan08,wiebe03},
all face-centered-cubic (FCC) AFs including type-I
systems (e.g. CeAs, CeSb, USb, NpBi)~\cite{furrer,bossy,hagen,jensen}, type-II systems (e.g. FeO, MnO, NiO, $\alpha$-MnS, CoO, EuTe, NiS$_{2}$)~\cite{yamamotoFCC}, type-III systems (e.g. Cd$_{1-x}$Mn$_x$Te for larger $x$) ~\cite{gieb86},
triangular stacked AF's~\cite{chubukov91triangle,chubukov94triangle,chern09triangle}, pyrochlore magnets~\cite{canals98Pyro,maged05Pyro}, kagome lattices~\cite{Fak08SL,harris92Kagome,subir92,chubu92Kagome,oleg08Kagome},
and fully frustrated cubic systems~\cite{diep85SC,viana,derrida80SC}.

The tendency of a magnetic system to support long range order is more pronounced in three dimensions (3D) than in two- or one- dimension. Recently, motivated by the results for the 2D lattices, some work has been done by analytical (non-linear spin wave theory) ~\cite{dattaBCC, dattaSC} and numerical techniques (exact diagonalization, and linked-cluster series expansions) ~\cite{oitmaa,Schmidt,viana} to understand the magnetic phase diagram of 3D quantum spin-$1/2$ Heisenberg AF on a body-centered-cubic (BCC) lattice and simple cubic (SC) lattice. There also exists a limited amount of work on the  effects of local anisotropy, four-spin exchange interactions, and biquadratic interaction of spin-$1/2$ Heisenberg AF on 3D lattices ~\cite{oguchi,katanin08FCC,ader02FCC,oja}.

\begin{figure}[h]
\centering
\subfigure[FCC Type IIA]{\includegraphics[width=1.5in]{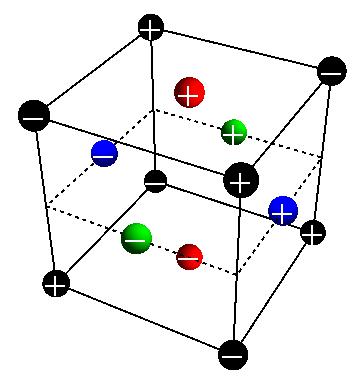}\label{subfig:Type2A}} \subfigure[FCC Type IIB]{\includegraphics[width=1.5in]{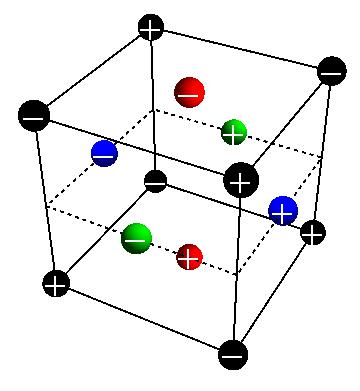}\label{subfig:Type2B}}
\caption{\label{fig:TypeAB} (Color online) Twofold structurally degenerate inequivalent Type IIA and Type IIB FCC structures. The four colors - \textcolor{black}{\bf black}, \textcolor{red}{{\bf red}}, \textcolor{blue}{\bf blue}, and \textcolor{green}{\bf green} represent the four sublattices respectively. The $+$ and $-$ denote the up and down collinear spin configurations respectively. Spin wave theory upto quartic order in the ratio of interaction strengths show that quantum fluctuations select the Type IIA structure as having the lower energy ~\cite{HarrisFCC}. We focus on this spin arrangement in this paper.}
\end{figure}

\begin{figure}[t]
\includegraphics[width=3.0in]{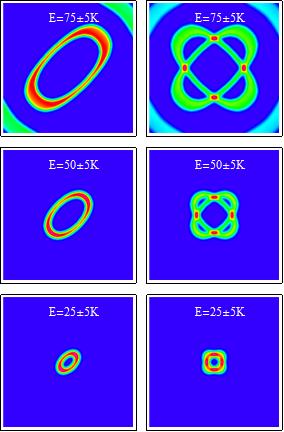}
\caption{\label{fig:MnOneutron} (Color online) Constant energy cuts of the inelastic neutron scattering pattern for MnO at {\bf q}=$(q_x,q_y,\pi)$. The x-axis and y-axis correspond to q$_{x}$  and q$_{y}$ respectively with the range (0, 2$\pi$). The left hand column is for an untwinned crystal and the right hand for a twinned crystal.  The parameter set and the scaled values of magnetic field, anisotropy, and bi-quadratic interaction used to compute the pattern are the same as the MnO spin wave dispersion plot.}
\end{figure}

\begin{figure}[t]
\includegraphics[width=3.0in]{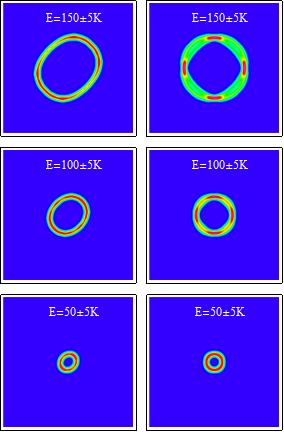}
\caption{\label{fig:CoOneutron} (Color online) Constant energy cuts of the inelastic neutron scattering pattern for CoO at {\bf q}=$(q_x,q_y,\pi)$. The x-axis and y-axis correspond to q$_{x}$  and q$_{y}$ respectively with the range (0, 2$\pi$). The left hand column is for an untwinned crystal and the right hand for a twinned crystal.  The parameter set and the scaled values of magnetic field, anisotropy, and bi-quadratic interaction used to compute the pattern are the same as the CoO spin wave dispersion plot.}
\end{figure}

\begin{figure}[t]
\centering
\includegraphics[width=3.0in]{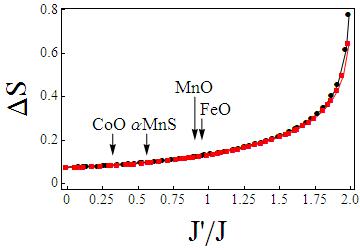}
\caption{\label{fig:deltaS}(Color online) Spin reduction, $\Delta S $, due to zero point quantum fluctuations for $J^{'}/J$ ratio in the absence of magnetic field, single-ion anisotropy, and bi-quadratic interaction.  Various Type II FCC antiferromagnets stated in Table~\ref{tab:FCCDatatable} are indicated on the graph.}
\end{figure}

In this paper, we study the effects of quantum fluctations in a 3D quantum AF on a FCC lattice using linear spin wave theory (LSWT). We choose the Type IIA FCC structure (see Fig.~\ref{subfig:Type2A}) which is proven to be stable from among the (initially) degenerate ground states of the FCC AF ~\cite{HarrisFCC,henley}. We then perform a LSWT calculation for spins in a canted basis about the Type IIA ground state and obtain the dispersion including the effects of external magnetic field, single-ion anisotropy, and biquadratic interaction (refer Eq.~\ref{eq:dispersion}). We calculate the expected finite frequency neutron scattering intensity for the FCC AF (see
Figs.~\ref{fig:MnOneutron} and ~\ref{fig:CoOneutron}). We also compute the effect of quantum fluctuations on the sublattice magnetization (see Fig.~\ref{fig:deltaS}).

The motivation for considering the Type IIA FCC lattice is twofold. First, as mentioned earlier, there are several experimentally relevant Type IIA FCC materials. Second, to the best of our knowledge a systematic theoretical study of neutron scattering and the effects of  magnetic field and single-ion anisotropy for the Type IIA FCC AF is missing. A knowledge of the neutron scattering pattern is crucial to determining information relevant for the quantum/classical dynamics of the system and to help understand the effects of frustration.The Type II FCC lattice had been an earlier topic of theoretical investigation where the authors studied the effect of frustration and quantum fluctuations ~\cite{HarrisFCC}. The Type I FCC AF has already been investigated in some detail theoretically ~\cite{ader01FCC,ader02FCC}.

This paper is organized as follows. In Section~\ref{sec:fccreview}
we begin with a brief description of the existing theoretical and
experimental understanding of the FCC AF system and the
properties of the lattice relevant to our calculations. In
Section~\ref{sec:lswt} we set-up the Hamiltonian and perform the
boson transformation for spins in a canted basis to obtain
the spin wave dispersion within LSWT. In
Section~\ref{subsec:dispersion} we discuss the effects of magnetic
field, single-ion anisotropy, and bi-quadratic interaction on the
dispersion relation. In Section~\ref{subsec:neutron} we present the
results of neutron scattering for MnO and CoO. In Section~ \ref{subsec:submag} we
show the effects of quantum fluctuations on the sublattice
magnetization. Finally, in Section~\ref{sec:discon} we
summarize the main results.

{\section{FCC review}\label{sec:fccreview}} The classical ground states of the FCC
AF have been investigated theoretically
~\cite{yamamotoFCC,oguchi}. Extending these studies to include
the effect of quantum fluctuations it was shown that the continuous
degeneracy of the classical ground states can be removed to favor a
collinear ground state ~\cite{shender82,henley}. In general there are 4 types or kinds of
collinear structure - Type I, Type IIA, Type IIB, Type III, and Type
IV. Quantum fluctuations are unable to remove the twofold structural
degeneracy between the inequivalent Type IIA and Type IIB structure
(see Fig.~\ref{fig:TypeAB} for spin arrangement). Classically
these two structures are stable for $|J^{'}|<2|J|$. As shown by
~\cite{HarrisFCC,henley} a spin wave theory up to order $(J^{'}/J)^4$ is
needed to lift the degeneracy to select the second kind of type A as
having the lower energy. In this paper we focus on this spin
arrangement. The spin wave gaps occur at the relative order of
$(J^{'}/J)^2$. The gap can be phenomenologically modeled by a
bi-quadratic interaction and microscopically justified through a spin
wave theory Hartree decoupling of quartic interaction terms ~\cite{HarrisFCC,henley}.

The Type II FCC AF system has also been investigated experimentally. The data in Table ~\ref{tab:FCCDatatable} documents the experimentally measured N\'{e}el transition temperature (T$_{N}$), nearest neighbor (nn) exchange interaction (J$^{'}$), next-nearest neighbor (nnn) exchange interaction (J), and the ratio of $J'/J$. The table shows that J $>$ J$^{'}$ for these materials. Under this condition the FCC lattice may be viewed as four interpenetrating SC AF sublattices in which the mean field on one sublattice due to any other vanishes. This fact forms the basis for writing our Hamiltonian, Eq.~\ref{eq:Hamiltonian}, in the sublattice formulation. In the next section, Sec.~\ref{sec:lswt}, we state the Hamiltonian and carry out the LSWT calculation in the canted spin basis about the FCC Type IIA AF ground state (see Fig.~\ref{subfig:Type2A}).

\begin{figure}[b]
\includegraphics[width=2.5in]{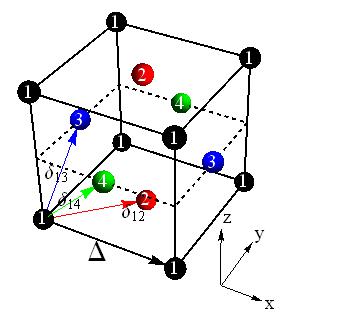}
\caption{\label{fig:fcc} (Color online) Four sublattice formulation of the AF FCC lattice. The sublattices are indicated by - \textcolor{black}{\bf black} (1), \textcolor{red}{{\bf red}} (2), \textcolor{blue}{\bf blue} (3), and \textcolor{green}{\bf green} (4) colors. The corresponding sublattice number is indicated in the parenthesis. The intra-sublattice nearest neighbor vector is given by ${\bf \Delta}$ and inter-sublattice vectors are given by $\Vec{\delta}_{12},\Vec{ \delta}_{13}$ and $\Vec{\delta}_{14}$. The choice of coordinate axis is as shown in the figure.}
\end{figure}
\begin{table}[t]
\caption{\label{tab:FCCDatatable} Transition temperature (T$_{N}$), nearest neighbor (J$^{'}$), next-nearest neighbor $ (J)$, and relative coupling strength $\gamma$ for Type II FCC antiferromagnets. The data has been compiled from Yamamoto and Nagamiya ~\cite{yamamotoFCC}.}
\begin{ruledtabular}
\begin{tabular}{ccccc}
   Material & T$_N$ [K (meV)] & J$^{'}$ [K (meV)] &J[K (meV)] & $\gamma$=J$^{'}$/J \\ \hline
   FeO & 198 (17.07) & 7.8 (0.67) & 8.2 (0.71) & 0.95\\
   MnO & 117 (10.08) & 5 (0.43) & 5.5 (0.47) &0.91\\
   NiO & 523 (45.09) & 50 (4.31)& 85 (7.33)& 0.59\\
   $\alpha$-MnS & 147 (12.7) & 3.5 (0.30) & 6.25 (0.54)&0.56\\
   CoO & 292 (25.2)& 6.9 (0.60) & 21.6 (1.86)&0.32\\
   EuTe & 9.76  (0.84)& 0.07 (0.006) & 0.21 (0.02)&0.33
    \end{tabular}
\end{ruledtabular}
\end{table}
{\section{Hamiltonian and LSWT}\label{sec:lswt}}
The model Hamiltonian that we study is given by,
\begin{eqnarray}
\mathcal{H}&=&J\sum_{\langle \alpha,i;\alpha,j\rangle}{{\bf S}_{\alpha,i}\cdot {\bf S}_{\alpha,j}}+J^{'}\sum_{\langle \alpha, i;\beta,j \rangle}{{\bf S}_{\alpha,i}\cdot {\bf S}_{\beta,j}}\nonumber \\&-&\lambda\sum_{ \alpha, i}( S^{z}_{\alpha,i})^{2}+g\mu_{B}H\sum_{\alpha, i}S^{y}_{\alpha, i}
\label{eq:Hamiltonian}
\end{eqnarray}
where {\bf S$_{\alpha,i}$} denotes the $i$th spin on a sublattice $\alpha$. The first term, with strength J is the interaction within the sublattices (which is a nn interaction based on the SC formulation). The second term with strength J$^{'}$ is the inter sublattice interaction. In Fig.~\ref{fig:fcc} we show the nn (${\bf \Delta}$) and nnn ($\Vec{\delta}_{12},\Vec{\delta}_{13}$ and $\Vec{\delta}_{14}$) vectors in the four sublattice SC formulation. The four colors - black (1), red (2), blue (3), and green (4) represent the four sublattices respectively. The corresponding sublattice number is indicated in the parenthesis. The third term is the single-ion anisotropy with strength $\lambda >0$, and the last term is the Zeeman energy due to the external magnetic field which has been applied along the negative y-direction ~\cite{syromaleyev}. The gyromagnetic ratio is given by the symbol g and the Bohr magneton by $\mu_{B}$.

In the presence of a magnetic field the sublattices are canted by an angle $\Theta$ and we apply the canted spin wave theory approach to calculate the LSWT dispersion ~\cite{zhitniku,zhitchern,syromaleyev,mourigal}. In this method the spin components are first represented in a rotating frame such that the local z$_{i}$-axis points in the direction of each magnetic sublattice. We then express the spin components in the rotated frame relative to the laboratory frame as,\begin{eqnarray}
&{\bf S}_{\alpha,i}&=\tilde{S^{x}}_{\alpha,i}{\bf \hat{x}}+\left(\tilde{S^{y}}_{\alpha,i}e^{i{\bf k_{o}}\cdot{\bf r}_{\alpha,i}}\cos\Theta+\tilde{S^{z}}_{\alpha,i}\sin\Theta\right){\bf \hat{y}}\nonumber\\
&+&\left(\tilde{S^{z}}_{\alpha,i}e^{i{\bf k_{o}}\cdot{\bf r}_{\alpha,i}}\cos\Theta-\tilde{S^{y}}_{\alpha,i}\sin\Theta\right){\bf \hat{y}}\label{eq:rotbasis}
\end{eqnarray} where {\bf k}$_{o}=(\frac{\pi}{a},\frac{\pi}{a},\frac{\pi}{a})$. The advantage of the rotated basis is that it allows us to perform the calculation using only one type of sublattice boson. In the second step of the process we obtain the boson representation (note that within LSWT there is no distinction between the Dyson-Maleev and the Holstein-Primakoff transformation) of the spins in the Hamiltonian by applying the transformation,
\begin{eqnarray}
\tilde{S}^{x}_{\alpha,i}&=&\sqrt{\frac{S}{2}}\left(a_{\alpha,i}+a^{\dag}_{\alpha,i}\right)\label{eq:sxdm}\\
\tilde{S}^{y}_{\alpha,i}&=&-i\sqrt{\frac{S}{2}}\left(a_{\alpha,i} - a^{\dag}_{\alpha,i}\right)\label{eq:sydm}\\
\tilde{S}^{z}_{\alpha,i}&=&S - a^{\dag}_{\alpha,i}a_{\alpha,i}\label{eq:szdm}
\end{eqnarray}
where S is the spin, $a^{\dag}_{\alpha, i}$ and $a_{\alpha,i}$  are the boson creation and annihilation operators for each site $i$ in a sublattice $\alpha$ respectively. Finally, we use the  transformation Eqs.~\ref{eq:sxdm},~\ref{eq:sydm}, and~\ref{eq:szdm}, to write the  Hamiltonian bilinear in the boson operators as,
\begin{eqnarray}
\mathcal{H}&=&-12NJS^{2}\cos2\Theta+4g\mu_{B}HNS\sin\Theta-4NS^{2}\lambda\cos^{2}\Theta\nonumber\\
&+&(-g\mu_{B}H\sin\Theta+2S\lambda\cos^{2}\Theta)\sum_{\alpha,i}a^{\dag}_{\alpha,i}a_{\alpha,i}\nonumber\\
&+&JS\sum_{\langle\alpha_{i},\alpha_{j}\rangle}\cos2\Theta(a^{\dag}_{\alpha,i}a_{\alpha,i}+a^{\dag}_{\alpha,j}a_{\alpha,j})\nonumber\\
&+&JS\sum_{\langle\alpha_{i},\alpha_{j}\rangle}\sin^{2}\Theta(a^{\dag}_{\alpha,i}a_{\alpha,j}+a_{\alpha,i}a^{\dag}_{\alpha,j})\nonumber\\
&+&JS\sum_{\langle\alpha_{i},\alpha_{j}\rangle}\cos^{2}\Theta(a^{\dag}_{\alpha,i}a^{\dag}_{\alpha,j}+a_{\alpha,i}a_{\alpha,j})\nonumber\\
&+&J^{'}S\sum_{\langle\alpha_{i},\beta_{j}\rangle}A^{'}_{\alpha_{i},\beta_{j}}(a^{\dag}_{\alpha,i}a_{\beta,j}+a_{\alpha,i}a^{\dag}_{\beta,j})\nonumber\\
&+&J^{'}S\sum_{\langle\alpha_{i},\beta_{j}\rangle}B^{'}_{\alpha_{i},\beta_{j}}(a^{\dag}_{\alpha,i}a^{\dag}_{\beta,j}+a_{\alpha,i}a_{\beta,j})
\label{eq:Hsite}
\end{eqnarray}
where $\Theta$ is the canting angle of the sublattice magnetization.  In the above derivation the linear terms from \~{S}$^{y}_{\alpha,i}$ do not contribute. We also ignore interaction terms of order $\lambda\sin^{2}\Theta$  which are small in the presence of weak field and anistropy. These terms could be relevant in the presence of strong magnetic field, but, we do not consider that analysis here. Furthermore, in the presence of strong magnetic field the present derivation will break down and a more careful analysis is required ~\cite{mourigal}. The canting angle, $\Theta$, can be obtained by minimizing the classical energy (the first three terms of Eq.~\ref{eq:Hsite}) to obtain,
\begin{equation}
\sin\Theta=-\frac{g\mu_{B}H}{12JS+2S\lambda}
\label{eq:cant}
\end{equation}
We now Fourier Transform the above Hamiltonian using the following definition $a^{\dag}_{\alpha,i}=\frac{1}{\sqrt{N}}\sum_{{\bf q}}a^{\dag}_{\alpha}({\bf q})e^{i{\bf q}\cdot {\bf r}_{i}}$ where N is the number of sites in each of the four SC AF sublattice, {\bf q} is summed over N values in the interval $-\pi < aq_{j}< \pi$ (j = x,y,z). The Fourier Transformed Hamiltonian may be written as a sum of the classical energy $\mathcal{E}_{cl}$, the intra (same) sublattice  interaction $\mathcal{H}_{intra}$, and inter (different) sublattice interaction $\mathcal{H}_{inter}$ as,
$\mathcal{H}=\mathcal{E}_{cl}+\mathcal{H}_{intra}+\mathcal{H}_{inter}$
where,\begin{equation}
\mathcal{E}_{cl}=-12NJS^{2}\cos2\Theta+4g\mu_{B}HNS\sin\Theta-4NS^{2}\lambda\cos^{2}\Theta
\label{eq:Ecl}
\end{equation}
\begin{eqnarray}
&\mathcal{H}_{intra}&= 6JS[\sum_{\alpha,{\bf q}}(1+\lambda/3J+\gamma_{\bf q}\sin^{2}\theta)a^{\dag}_{\alpha}({\bf q})a_{\alpha}({\bf q})\nonumber\\
&+& \frac{1}{2}\gamma_{\bf q}\cos^{2}\theta\sum_{\alpha,{\bf q}}(a^{\dag}_{\alpha}({\bf q})a^{\dag}_{\alpha}({-\bf q})+a_{\alpha}({\bf q})a_{\alpha}({-\bf q})]]\nonumber\\
\label{eq:hintra}
\end{eqnarray}
and $\gamma_{\bf q}$ is given by $\gamma_{\bf q}=\frac{1}{6}\sum_{\bf \Delta}e^{i{\bf q}\cdot{\bf \Delta}}$
with ${\bf \Delta}$ as the nn neighbor vector within a sublattice  (see Fig.~\ref{fig:fcc}). Finally,
\begin{eqnarray}
&\mathcal{H}_{inter}&= J^{'}S[\sum_{\alpha,\beta,{\bf q}}A^{'}_{\alpha,\beta}({\bf q})[a^{\dag}_{\alpha}({\bf q})a_{\beta}({\bf q})+a_{\alpha}({\bf q})a^{\dag}_{\beta}({\bf q})]\nonumber\\
&+& B^{'}_{\alpha,\beta}({\bf q})\sum_{\alpha,\beta,{\bf q}}(a^{\dag}_{\alpha}({\bf q})a^{\dag}_{\beta}({-\bf q})+a_{\alpha}({\bf q})a_{\beta}({-\bf q})]]\nonumber\\
\label{eq:hinter}
\end{eqnarray}
The A$^{'}_{\alpha,\beta}({\bf q})$ and B$^{'}_{\alpha,\beta}({\bf q})$ coefficients are given by
\begin{eqnarray}
A^{'}_{\alpha,\beta}{\bf (q)}&=&\frac{1}{4}\sum_{{\Vec{\delta}}_{\alpha,\beta}}\left[1+\left(e^{i{\bf k_{o}}\cdot{\Vec{\delta}}_{\alpha,\beta}}\cos^{2}\Theta+\sin^{2}\Theta\right)\right]e^{-i{\bf q}\cdot{\Vec{\delta}}_{\alpha,\beta}}\nonumber\\\\
B^{'}_{\alpha,\beta}{\bf (q)}&=&\frac{1}{4}\sum_{{\Vec{\delta}}_{\alpha,\beta}}\left[1-\left(e^{i{\bf k_{o}}\cdot{\Vec{\delta}}_{\alpha,\beta}}\cos^{2}\Theta+\sin^{2}\Theta\right)\right]e^{-i{\bf q}\cdot{\Vec{\delta}}_{\alpha,\beta}}\nonumber\\
\label{eq:NewAB}
\end{eqnarray}
where ${\Vec{\delta}_{\alpha,\beta}}$ is summed over the four first-neighbor vectors which connect sublattices $\alpha$ and $\beta$ (see Fig.~\ref{fig:fcc}). With the above definitions the bilinear Hamiltonian, Eq.~\ref{eq:Hsite}, can be written as follows,
\begin{equation}
\mathcal{H}=\mathcal{E}_{cl}+\frac{1}{2}\sum_{{\bf q}}{\bf X}^{\dag}({\bf q}){\bf M}({\bf q}){\bf X}({\bf q})
\label{eq:Hdiag}
\end{equation}
with the following defintions,
\begin{eqnarray}
{\bf X}({\bf q})=\left(\begin{tabular}{c}
  {\bf V}({\bf q}) \\
{\bf V}$^{\dag}$(-{\bf q})\\
\end{tabular}
\right)\\
{\bf V}({\bf q})=\left(\begin{tabular}{c}
a$_{1}$({\bf q}) \\
a$_{2}$({\bf q}) \\
a$_{3}$({\bf q}) \\
a$_{4}$({\bf q}) \\
\end{tabular}\right)\\
{\bf M}({\bf q})=\left(\begin{tabular}{c c}
{H}$_{1}$({\bf q})  &{H}$_{2}$({\bf q})\\
{H}$_{2}$({\bf q}) &{H}$_{1}$({\bf q})
\end{tabular}\right)
\end{eqnarray}
and H$_{1}$({\bf q}) and H$_{2}$({\bf q}) are defined as, 
\begin{eqnarray}
H_{1}({\bf q})&=&6JS\{(1+\lambda/3J+\gamma_{\bf q}\sin^{2}\theta)+[J^{'}/(3J)]{A}^{'}({\bf q})\}\nonumber\\
\label{eq:eqhone}
\end{eqnarray}
\begin{eqnarray}
H_{2}({\bf q})&=&6JS\{\gamma_{\bf q}\cos^{2}\theta+[J^{'}/(3J)]{B}^{'}({\bf q})\}
\label{eq:eqhtwo}
\end{eqnarray}
Now in the above Hamiltonian one can include the biquadratic interaction defined by ~\cite{HarrisFCC},
\begin{equation}
\Delta\mathcal{H}_{Q}=-\frac{1}{2}Q\sum_{i,j}\Delta_{ij}[{\bf S}_{i}\cdot{\bf S}_{j}]^{2}/S^{3}
\label{eq:biquad}
\end{equation}
where Q is the strength of the biquadratic interaction, $\Delta_{ij}$ is unity if spins i and j are nn and is zero otherwise. Following the same strategy as outlined before we have the following \emph{new definitions} of H$_{1}({\bf q})$ and H$_{2}({\bf q})$ in the presence of biquadratic interaction,
\begin{eqnarray}
H_{1}({\bf q})&=& 6JS\left[1+\frac{\lambda}{3J}+\gamma_{{\bf q}}\sin^{2}\Theta+\frac{2Q}{JS}\right]\\
&+&6JS\left[2j - \frac{Q}{3JS}\right]h_{1}({\bf q})+6JS\left[2j\sin^2\Theta\right]h_{2}({\bf q})\nonumber\\
H_{2}({\bf q})&=&6JS\gamma_{{\bf q}}\cos^{2}\Theta +6JS\left[2j\cos^{2}\Theta+\frac{Q}{3JS}\right]h_{2}({\bf q})\nonumber\\
\label{eq:NewH1H2}
\end{eqnarray}
where h$_{1}$({\bf q}) and  h$_{2}$({\bf q}) are now given by, \begin{eqnarray}
h_{1}({\bf q})&=&\cos[a(q_x-q_y)/2]
+\cos[a(q_y-q_z)/2]\nonumber\\&+&\cos[a(q_z-q_x)/2]\\
h_{2}({\bf q}) &=&\cos[a(q_x+q_y)/2]+\cos[a(q_y+q_z)/2]\nonumber\\
&+&\cos[a(q_z+q_x)/2]
\end{eqnarray} with  j=$J^{'}/6J$.
We choose the fourth eigenvalues h$_{1}$({\bf q}) and h$_{2}$ ({\bf q}) because H$_{1}$({\bf q}) and H$_{2}$({\bf q}) commute and can be simultaneously diagonalized for the
Type IIA structure (see appendix A of Ref. [39] for more
details). The other three branches can be obtained by "folding" the single branch spectrum ~\cite{HarrisFCC}. The neutron scattering is nonzero only for the single mode appearing in this basis. Hence the choice of this representation is convenient to perform the neutron scattering calculation. The LSWT dispersion is given by,
\begin{equation}
\omega({\bf q})=\sqrt{[H_{1}({\bf q}) - H_{2}({\bf q})][H_{1}({\bf q}) + H_{2}({\bf q})]}
\label{eq:dispersion}
\end{equation}
\begin{figure}[t]
\centering
\subfigure[MnO]{\includegraphics[width=3.5in]{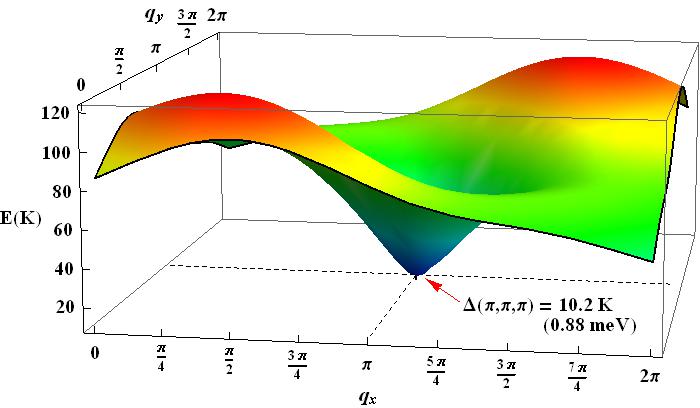}\label{subfig:MnO}}
\subfigure[CoO]{\includegraphics[width=3.5in]{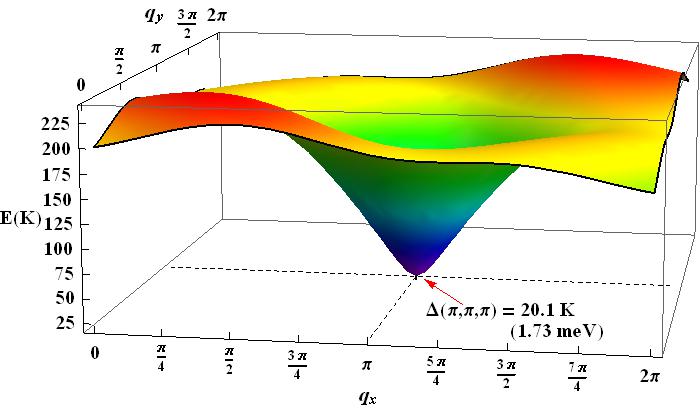}\label{subfig:CoO}}
\caption{\label{fig:SWD}(Color online) Linear spin-wave theory dispersion for (a) MnO and (b) CoO. The x-axis and y-axis correspond to wave-vectors q$_{x}$  and q$_{y}$ respectively with the range (0,2$\pi$). The z-axis corresponds to energy in units of kelvin. The value of q$_z=\pi$. We compute the dispersions for the parameter set (S=5/2, J$^{'}$=5 K, J=5.5 K) for MnO and (S=3/2, J$^{'}$=6.9 K, J=21.6  K) with the ratios $\frac{g\mu_{B}H}{6J}$=0.05, $\frac{\lambda}{6J}=0.002$, and $\frac{Q}{J}=0.01$. For this choice of parameters the energy gap at the antiferromagnetic $(\pi,\pi,\pi)$ point $\Delta(\pi,\pi,\pi)$, indicated by a red arrow on the plot, is 10.2 K (0.88 meV) for MnO and 20.1 K (1.73 meV) for CoO. The dashed lines in the q$_x$-q$_y$ plane are a guide for the eye to locate the $(\pi,\pi,\pi)$ point. The spin of the compound is given by S, J$^{'}$ is the inter sublattice interaction, J is the interaction within same sublattice, $\lambda$ is the anisotropy parameter, Q is the bi-quadratic interaction, and H is the external magnetic field.}
\end{figure}
 {\section{Results}\label{sec:results}}
{\subsection{Spin wave dispersion - effects of magnetic field,
anisotropy, and bi-quadratic interaction}\label{subsec:dispersion}}
The spin wave dispersion for MnO and CoO is displayed in Fig.~\ref{fig:SWD} for (q$_x$, q$_y$, $\pi$). The range for q$_{x}$ and q$_{y}$ is 0 to 2$\pi$ on the dispersion plot. The energy is measured in units of kelvin. We use
Eq.~\ref{eq:dispersion} to compute the energy profile for the parameter set (S, J$^{'}$, J) with the ratios $\frac{g\mu_{B}H}{6J}$=0.05, $\frac{\lambda}{6J}=0.002$, and $\frac{Q}{J}=0.01$. The spin of the compound is given by S, J$^{'}$ is the inter sublattice interaction, J is the interaction within same sublattice, $\lambda$ is the anisotropy parameter, Q is the bi-quadratic interaction, and H is the external magnetic field. From Table \ref{tab:FCCDatatable} and 
Ref.~\onlinecite{HarrisFCC} we take the parameter set (5/2, 5 K (0.43 meV), 5.5 K(0.47 meV)) for MnO and (3/2, 6.9 K (0.60 meV), 21.6 K(1.86 meV)) for CoO. We find that for the given choice of parameters the AF gap at the $(\pi,\pi,\pi)$-point is 10.2 K (0.88 meV) for MnO and 20.1 K (1.73 meV) for CoO. The two compounds also have a different band top - 134 K (11.6 meV) for MnO and 239 K (20.6 meV) for CoO. The dispersion also shows that the energy range is different and there is an overall qualitative difference in the curvature of the dispersion.

\begin{figure}[t]
\centering
\includegraphics[width=3.5in]{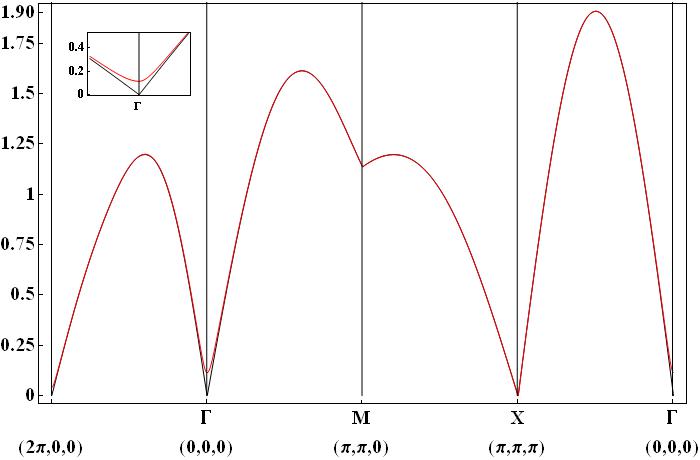}
\caption{\label{fig:magfield} (Color online) MnO spin wave dispersion, $\omega$({\bf q})/15J, for zero external magnetic field (solid black line) and magnetic field $\frac{g\mu_{B}H}{6J}$=0.15 (solid red line). The field ratio is chosen to emphasize the effect of magnetic field on the spin wave dispersion. An energy gap of 9.45 K (0.81 meV) opens up at the $\Gamma$-point as displayed in the inset. Smaller values of the field produces a smaller gap. The Brillouin zone is traversed along (2$\pi$,0,0), $\Gamma$, X, M, and then back to $\Gamma$.The inset is displayed for $(\frac{\pi}{8},0,0) \rightarrow \Gamma \rightarrow (\frac{\pi}{8},\frac{\pi}{8},0)$.}
\end{figure}

\begin{figure}[t]
\centering
\includegraphics[width=3.5in]{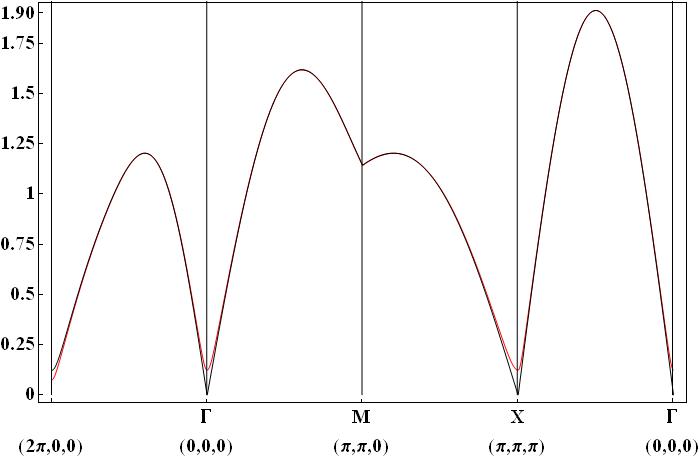}
\caption{\label{fig:andis} (Color online) MnO spin wave dispersion, $\omega$({\bf q})/15J, with single-ion anisotropy parameter $\frac{\lambda}{6J}=0.002$ (solid red line) and bi-quadratic interaction parameter $\frac{Q}{J}=0.01$ (solid black line). In the presence of single-ion anisotropy only, multiple energy gaps are developed at the high symmetry points -  10.2 K (0.87 meV) at $\Gamma$ \& X and 6.2 K (0.53 meV) at $(2\pi,0,0)$. In the presence of bi-quadratic interaction only, a single energy gap of 10.1 K (0.86 meV) opens up at the $(2\pi, 0, 0)$-point in confirmation with ~\cite{HarrisFCC}. The Brillouin zone is traversed along (2$\pi$,0,0), $\Gamma$, X, M, and then back to $\Gamma$.}
\end{figure}

In Figs.~\ref{fig:magfield} and ~\ref{fig:andis} we display the dispersion curves for MnO, $\omega$({\bf q})/15J, for different cuts along the Brillouin zone (BZ) - (2$\pi$, 0, 0), $\Gamma$, M, X, and then back to $\Gamma$. From Fig.~\ref{fig:magfield} we observe that in the absence of a magnetic field (solid black line) there are no gaps in the BZ. But, with the inclusion of a field (solid red line) an energy gap of 9.45 K (0.81 meV) opens up at the $\Gamma$-point as shown in the inset of  Fig.~\ref{fig:magfield}. The value of this gap will increase if the strength of the field is increased (for reasons mentioned in Section \ref{sec:lswt} a very large value of the magnetic field should not be used). Gaps can also be created due to the presence of single-ion anisotropy (solid black line) and bi-quadratic interaction (solid red line) as shown in Fig.~\ref{fig:andis}. With single-ion anisotropy present only, a gap of 10.2 K (0.87 meV) opens up at the $\Gamma$ \& X point and 6.2 K (0.53 meV) at the (2$\pi$, 0, 0) point. However, with the biquadratic interaction a gap of 10.1 K (0.86 meV) is present \emph{only} at the (2$\pi$, 0, 0) point and this fact is in agreement with the work of Yildirim \emph{et.al.}~\cite{HarrisFCC}. The CoO dispersion cuts along the BZ (not shown here) also display a qualitatively similar behavior. 
\begin{figure}[t]
\centering
\includegraphics[width=3.0in]{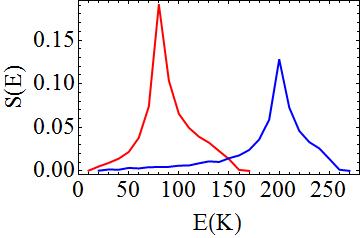}
\caption{\label{fig:swdisp} (Color online) Integrated structure factor S(E) for the full Brillouin zone. The computations were done for the same parameter set as in the neutron intensity calculation. The red line represents the structure factor for MnO and the blue line for CoO. The energy is in units of kelvin.}
\end{figure}
{\subsection{Inelastic neutron Scattering}\label{subsec:neutron}}
Neutron scattering is a very useful tool to detect magnetic order in
crystals ~\cite{boothroyd,tranquada,dai,tranaip}. Neutron
diffraction (elastic scattering) can be used to determine the spin
structure and magnetic moments. Inelastic neutron scattering can be
used to study spin dynamics including spin waves
~\cite{ewcneutron,yaoPRL,yaoPRB}. The technique has been used
successfully in magnetic materials, high T$_C$ superconductors, and
manganites.

The neutron scattering cross section is proportional to the dynamic
structure factor S({\bf q}, $\omega$) ~\cite{ewings}. In the linear
spin wave approximation the transverse parts contribute to the
structure factor and by symmetry we have,
\begin{eqnarray}
\frac{S^{\alpha\alpha}({\bf q},\omega)}{g^{2}\mu^{2}_{B}S_{eff}}
=\frac{H_{1}({\bf q})-H_{2}({\bf q})}{2\omega({\bf q})}[n(\omega)+1]\delta(\omega - \omega({\bf q}))\nonumber\\
\end{eqnarray} where $\alpha$=x,y and and n($\omega$) is the Bose occupation
factor. In Fig.~\ref{fig:MnOneutron} (MnO) and Fig.~\ref{fig:CoOneutron} (CoO) we show the expected neutron scattering intensity for constant energy cuts in {\bf q} space. The calculated spectra are predictions from LSWT. In either figure, the left hand column shows the expected neutron scattering intensity from a single domain of the magnetic order (untwinned cyrstal) and the right hand column shows the expected neutron scattering intensity from domains with both orientations of the magnetic order (twinned cyrstal). In real materials twinning occurs due to a finite correlation length, local disordered pinning, or crystal twinning.

We use the value of spin for the compound (S), J$^{'}$, J, H, $\lambda$, and Q. For MnO we have (5/2, 5 K (0.43 meV), 5.5 K (0.47 meV)) and for CoO (3/2, 6.9 K (0.60 meV), 21.6 K (1.86 meV)).
We choose $\frac{g\mu_{B}H}{6J}$=0.05, $\frac{\lambda}{6J}=0.002$, and
$\frac{Q}{J}=0.01$ as the scaled values of the magnetic field, single-ion anisotropy, and bi-quadratic interaction. The spectra are computed at {\bf q}=$(q_x,q_y,\pi)$ for both MnO and
CoO. The x-axis and y-axis correspond to q$_{x}$  and q$_{y}$
respectively with the range (0, 2$\pi$). The value of Q obtained
from Ref.~\onlinecite{HarrisFCC} is of the order of 0.001 meV. Such
a small value has negligible effect on the dispersion and the
neutron scattering plot. Therefore to highlight the effect of Q we
choose a slightly higher value. The choice of $\lambda$ is typical for AFs ~\cite{yao}.

For the untwinned case, at low energies the strongest diffraction peaks are centered around the ($\pi$,$\pi$) point and has an elliptical shape for MnO and a more circular shape for CoO.  As the energy is increased the elliptical neutron pattern for MnO is stretched out and the circular pattern for CoO increases in size. Simultaneously the neutron intensity starts to non uniformly concentrate along the edges of the ellipse connecting the (0,0) and the (2$\pi$, 2$\pi$) line for MnO. For the CoO case the intensity starts to spread uniformly along the circular ring. At even higher energies the MnO spectra acquires a flattened elliptical shape. The CoO spectra on the other hand becomes a distorted circular shape. The neutron scattering patterns are determined mainly by the ratio of J$^{'}$/J. For the twinned case the intensity is located at particular points. As the energy is increased the high intensity patches simply grow in size.

The intensity patterns for the MnO and CoO case are different at high energy. This fact is reflected in the integrated structure factor which is given by,
\begin{equation}
S^{\alpha\alpha}(\omega)=\int\int\int_{BZ}dk_{x}dk_{y}dk_{z}S^{\alpha\alpha}({\bf q},\omega)\delta(\omega-\omega({\bf q}))
\end{equation}
where $\alpha=x,y$ and BZ means integrate over the full magnetic
Brillouin zone. Numerical results for the structure factor are presented in
Fig.~\ref{fig:swdisp}. The energy scale at which the peaks appear
are different. The MnO peaks at a lower energy $\approx$
80 K ( 6.9 meV) while the CoO at a higher energy $\approx$ 200 K (17.2 meV).
{\subsection{Sublattice Magnetization}\label{subsec:submag}}
Fig.~\ref{fig:deltaS} shows the spin reduction, $\Delta S$, due to zero point quantum fluctuations, \begin{equation}
\Delta S=-\frac{1}{2}+\frac{1}{8N}\sum_{{\bf q}}\frac{H_{1}({\bf q})}{\omega({\bf q})}
\label{eq:deltaS}
\end{equation}
For spin suppression the various Type II FCC AFs are also indicated
on the graph. The black line in Fig.~\ref{fig:deltaS} is the quantum
fluctuation in the absence of magnetic field, single-ion anisotropy,
and biquadratic interaction. The red line shows the effect of
single-ion anisotropy on quantum fluctuations only. With the anisotropy value that we choose there is not much of a change from when it is absent. Both curves lie on top of each other. On the
figure the various Type II FCC AFs are indicated.

{\section{Summary and Conclusion}\label{sec:discon}} In this paper we compute the
LSWT dispersion for a FCC AF in the presence of magnetic field,
single-ion anisotropy, and biquadratic interaction. We carry out the spin wave theory computation about the Type IIA FCC structure which is known to be stable. We highlight the effects of magnetic field, single-ion anisotropy, and
biquadratic interaction on the FCC dispersion and the energy gaps that can be created. Using MnO and CoO as typical Type II FCC materials we compute the predicted inelastic neutron scattering pattern. The two predicted patterns differ at high energies and in the ellipticity of their plots which is controlled by the ratio of J$^{'}$/J. The effects of quantum fluctations on sublattice magnetization at zero temperature is also explored for various ratios of nn and nnn interactions.
\begin{acknowledgments}
T.D. acknowledges the invitation, kind hospitality, and research
funding support from Sun Yat-sen University and Fundamental Research
Funds for the Central Universities. T. D. also thanks Augusta State
University Katherine Reese Pamplin College of Arts and Sciences for
partial research funding support. D. X. Y. is supported by the
NSFC-11074310, Sun Yat-sen University, and Fundamental Research
Funds for the Central Universities.
\end{acknowledgments}
\bibliography{fccref}
\end{document}